\documentclass[pra, reprint,superscriptaddress,longbibliography]{revtex4-2}
\usepackage{amsmath}
\usepackage{amsfonts}
\usepackage{graphicx}
 \usepackage{physics}
 \usepackage[usenames]{color}

 \def\be{\begin{equation}} \def\ee{\end{equation}}
\def\bea{\begin{eqnarray}} \def\eea{\end{eqnarray}}

\begin{document}

\date{\today}

\title{Route to Extend the Lifetime of a Discrete Time Crystal in a Finite Spin Chain Without Disorder}
\author{Sayan Choudhury}
\affiliation{Department of Physics and Astronomy, University of Pittsburgh, Pittsburgh, PA 15260, USA; sayan.choudhury@pitt.edu}
\email{sayan.choudhury@pitt.edu}

\begin{abstract}
Periodically driven (Floquet) systems are described by time dependent Hamiltonians that possess discrete time translation symmetry. The spontaneous breaking of this symmetry leads to the emergence of a novel non-equilibrium phase of matter - the Discrete Time Crystal (DTC). In this paper, we propose a scheme to extend the lifetime of a DTC in a paradigmatic model - a translation invariant Ising spin chain with nearest-neighbor interaction $J$, subjected to a periodic kick by a transverse magnetic field with frequency $\frac{2 \pi}{T}$. This system exhibits the hallmark signature of a DTC - persistent subharmonic oscillations with frequency $\frac{\pi}{T}$ - for a wide parameter regime. Employing both analytical arguments as well as exact diagonalization calculations, we demonstrate that the lifetime of the DTC is maximized, when the interaction strength is tuned to an optimal value, $JT = \pi$. Our proposal essentially relies on an interaction induced quantum interference mechanism that suppresses the creation of excitations, and thereby enhances the DTC lifetime.  Intriguingly, we find that the period doubling oscillations can last eternally in even size systems. This anomalously long lifetime can be attributed to a time reflection symmetry that emerges at $JT=\pi$. Our work provides a promising avenue for realizing a robust DTC in various quantum emulator platforms.
\end{abstract}

\maketitle
\section{Introduction}
The classification of phases of matter on the basis of symmetries and symmetry breaking forms one of the cornerstones of modern statistical physics \cite{landau2013course}. Crystals represent a ubiquitous example of this paradigm, where inter-atomic interactions leads to the breaking of spatial translation symmetry \cite{beekman2019introduction}. In a seminal paper in 2012, Frank Wilczek extended the concept of symmetry breaking to the time domain \cite{wilczek2012prl}. In particular, he postulated that the ground state of attractively interacting bosons on a Aharonov–Bohm ring can exhibit periodic oscillations in time; this novel phase of matter was dubbed a ``Quantum Time Crystal", since it breaks time translation symmetry. Wilczek's bold proposal generated considerable excitement and debate \cite{li2012prl,bruno2013prl,thomas2013physics,nozieres2013epl}. Eventually however, the existence of quantum time crystals in equilibrium was ruled out by a no-go theorem \cite{watanbe2015prl}. This crucial insight has prompted several researchers to explore the possibility of realizing time crystals in non-equilibrium quantum systems \cite{sachapra2015,sondhi2016prl,sondhi2016prb,yao2017prl,nayak2016prl,nayak2017prx,yao2019arxiv,sheng2017prb,jaksch2019nature,fazio2018prl,jaksch2019prl,basak2019arxiv,lewenstein2019njp,alicea2019arxiv,hemmerich2019pra,mathey2019pra,apal2019arxiv,gong2019prb}. These efforts have been immensely fruitful and led to the development of an active area of research \cite{sacha2018review,nayak2019review,khemani2019review,sacha2020book,watanabe2020jsp,kozin2019prl,wright2019prl,khemani2020arxiv,sacha2020arxiv,sacha2020prr,kessler2020continuous,giergiel2020creating,yarloo2020homogeneous}.  \\

A particularly profound development in this field has been the discovery of discrete time crystals in periodically driven (Floquet) quantum many-body systems \cite{sondhi2016prl,sondhi2016prb,yao2017prl,nayak2016prl}. A Floquet system is described by a time-periodic Hamiltonian, $H(t)$ where $H(t+T) = H(t)$. A discrete time crystal (DTC) is an out-of-equilibrium phase of matter that breaks this time translation symmetry, and consequently shows a stable sub-harmonic response of physical observables. In particular, the DTC phase is characterized by the existence of a class of observables $O$ and initial states, $\vert \psi \rangle $, such that $\langle  \psi \vert O(t) \vert \psi \rangle   \ne \langle  \psi \vert O(t+T) \vert \psi \rangle$. Furthermore, in order to qualify as a genuine non-equilibrium phase of matter, these sub-harmonic oscillations must persist at long times ($t/T \gg 1$), without fine tuning the Hamiltonian parameters \cite{liu2018prl,fazio2017prb}. Through a number of theoretical and experimental studies, the existence of the DTC phase has now been firmly established \cite{monroe2017nature,lukin2017nature,ueda2018prl,lazarides2019arxiv,gambettaprl2019,yao2020nature,kawakami2018prl,pizzi2019prl,pizzi2020arxiv,zhu2019njp,fazio2017prb,liu2018prl,fazio2019prb,gambetta2019pre,fazio2020prr,barrett2018prl,mahesh2018prl}.\\

\begin{figure}
\includegraphics[scale=0.225]{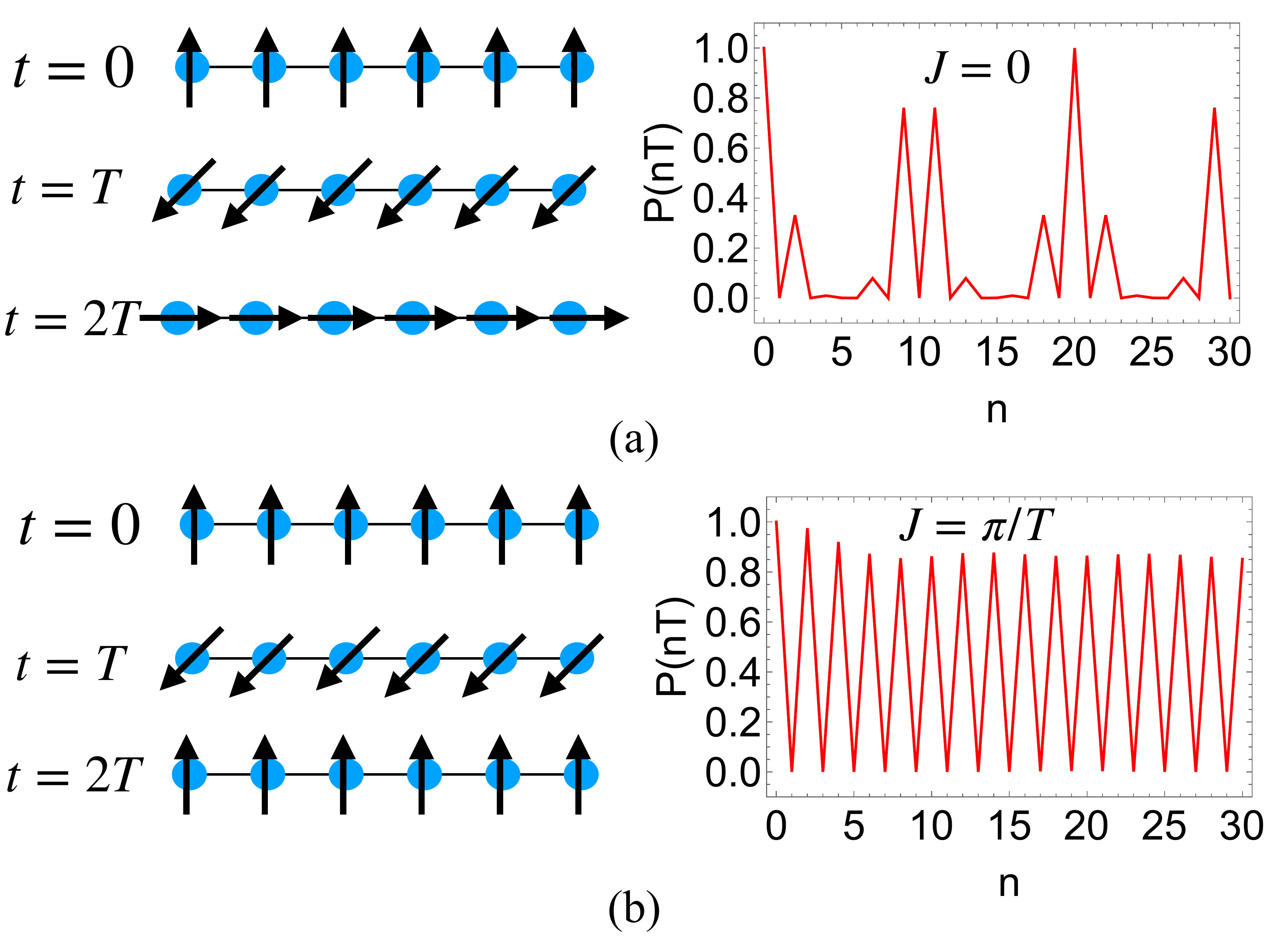}
\caption{(a)  The return probability, $P(nT)$ (defined in Eq.~({\ref{le}})) when $J=0$. When the $\pi$ pulse is imperfect, then there is no oscillation of the order       parameter and hence no time crystal. (b) The return probability, $P(nT)$ when $JT = \pi$ for the nearest-neighbor Ising model. This example clearly shows that persistent time crystal order can be established in a disorder free spin chain by tuning the interaction strength. }
\label{fig1}
\end{figure}

The first realization of a DTC crucially relied on the presence of many-body localization (MBL) \cite{monroe2017nature}. This is because Floquet many-body localized systems do not absorb energy from the drive, and consequently evades thermalization. This leads to persistent oscillations of physical observables with a characteristic frequency, $\omega \ne \frac{2 \pi}{T}$ at long times. However, the requirement of MBL is very restrictive, thereby making the realization of a DTC in large systems extremely challenging \cite{vsuntajs2020quantum,kiefer2021slow}. Furthermore, MBL can lead to long-lived transient dynamics, thereby making it difficult to access the long-time behavior of the system in current experiments \cite{monroe2021arxiv}. These issues have prompted researchers to examine other routes for realizing a robust DTC. \\

Intriguingly, recent research efforts have shown that it is possible to realize a robust DTC in the absence of disorder \cite{fazio2017prb,liu2018prl,fazio2019prb,gambetta2019pre,fazio2020prr,barrett2018prl,mahesh2018prl}.  This surprising observation implies the presence of mechanisms other than MBL that can prevent Floquet systems from heating up. Such mechanisms may also be useful for stabilizing other non-equilibrium phases of matter. To explore this issue in a concrete example, we analyze the conditions necessary for creating a time crystal in a periodically driven Ising spin chain. Yu {\it et al.} have already demonstrated that this system can exhibit discrete time-translation-symmetry-breaking (TTSB) and argued that the DTC order is stabilized by high frequency driving  \cite{jaksch2019pra}. High frequency driving can also been used to realize prethermal time crystals \cite{nayak2017prx,yao2019arxiv,sheng2017prb,monroe2021arxiv}. Unfortunately, when the driving frequency is very high, the DTC order can be destroyed due to dissipative coupling to higher bands \cite{weinberg2015pra}. This necessitates the search for other pathways to stabilize time crystals.\\

It has been recently demonstrated that Floquet phases of matter can be stabilized by a many-body quantum interference, thereby circumventing the problems associated with high frequency driving \cite{chin2019nature,zhai2019arxiv,zhou2019arxiv}. In particular, Lyu {\it et al.} have shown that it is possible to realize an eternal DTC in a periodically driven, even-size, infinite range interacting Ising spin chain, by appropriately tuning the interaction strength \cite{zhou2019arxiv}. This naturally raises the question of whether a similar scheme can be used to stabilize a DTC in a kicked short range Ising model. We answer this question affirmatively, and derive the optimal interaction strength that maximizes the DTC lifetime in a finite chain. An important aspect of our scheme is that the optimal interaction parameter maximizes the lifetime for both even and odd size chains, while the protocol presented in  ref.~\cite{zhou2019arxiv} is only applicable for even size chains. It is interesting to note that for this optimal interaction strength $J$, strong disorder leads to thermalization. Our results provide a novel route for extending the lifetime of DTCs in translation invariant systems, and can guide the experimental realizations of DTCs \cite{zinner2019prb,barnes2019prb,rossini2020pra,barnes2019arxiv}.\\

Before proceeding further, we note that the phenomenology of DTCs is a closely related to the dynamical Casimir effect, where periodic modulation of boundaries or material properties of a system leads to a parametric amplification of initial vacuum fluctuations \cite{dodonov2010current,dodonov2020fifty}. A particularly striking example of this phenomenon is the temporal Bragg diffraction, where a periodic modulation of the effective charge of neutral atoms leads to phonon backscattering in time. In this case, the modulation frequency $\omega$ is related to the initial phonon frequency $\omega_c$ as: $\omega = (n \pm 1) \omega_c/2$, where $n$ is an integer \cite{dodonov2014dynamical,mendonca2017time}. This is analogous to the discrete TTSB that characterizes DTCs.\\

This paper is organized as follows. In section 2, we introduce our model and discuss the conditions under which the system behaves like a discrete time crystal. In section 3, we describe our scheme to extend  the DTC lifetime by tuning the Ising interaction. In particular, we analytically derive the optimal interaction strength that maximizes the lifetime of the DTC. Furthermore, we establish the validity of our analytical results by numerically analyzing the dependence of the lifetime on the interaction strength as well as the system size. Thus, our calculations provide a general framework to realize time crystals in disorder free spin chains. Finally, we discuss potential realizations of our model. We conclude with a brief summary of our results in section 4. 

\section{Model}

\begin{figure}
\includegraphics[scale=0.225]{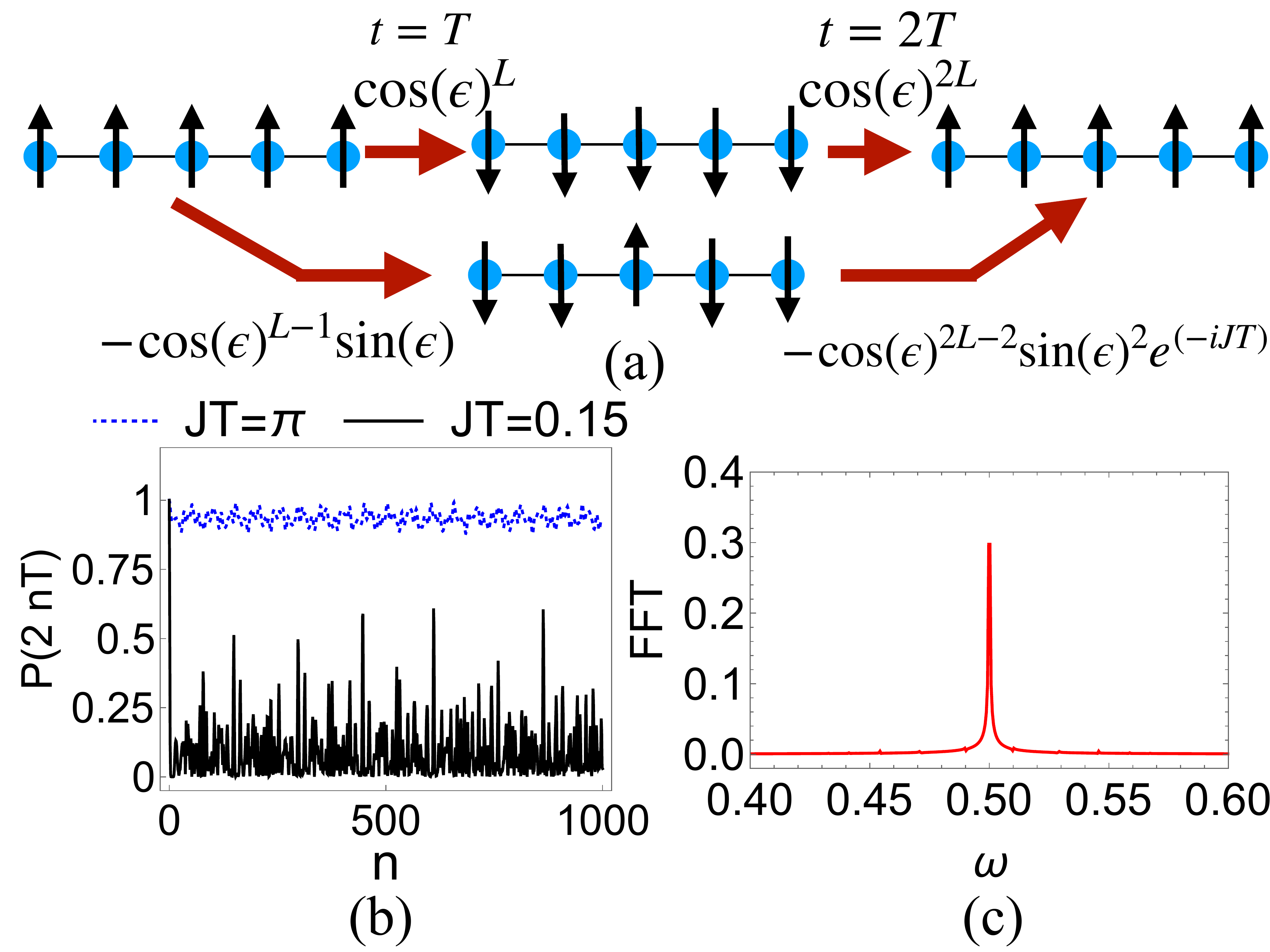}
\caption{
(a) Illustration of the state of the spin chain after two imperfect $\pi$ pulses, starting from a fully spin-polarized state. After the first pulse the spin chain wavefunction primarily comprises two classes of states: A fully spin-polarized state, and $L$ one-magnon states (i.e. a state with $L-1$ spin-downs and one spin-up). When $JT = \pi $, the two pathways interfere constructively, leading to a revival of the initial state. (b) The stroboscopic return probability $P(2nT)$ for two different values of $JT$, when $\epsilon = 0.07 \pi$. Persistent oscillations can be observed when $JT=\pi$. (c) The Fourier spectrum of the return probability shows a robust peak at $\omega_0/2$, where $\omega_0 = \frac{2 \pi}{T}$, when $JT = \pi$.
}
\label{fig2}
\end{figure}

We study the time evolution of a 1D spin-chain described by the following Floquet Hamiltonian:
\be
H =  \sum_{i}\left (J S^{z}_i S^{z}_{i+1}, + (\pi-2 \epsilon) S^{x}_i, \delta(t-nT) \right), 
\label{model}
\ee
where $J$ is the nearest-neighbor Ising interaction, $\vec{S}^i = \frac{\hbar}{2} \vec{\sigma}^i$ and $\vec{\sigma}^i$ are the Pauli spin matrices.\\
 
Setting $\hbar = 1$, we find the time-evolution operator for one Floquet period to be:  
\be
U(T) = e^{(-i H T)} \equiv e^{(-i \frac{J T}{4} \sum_i \sigma^{z}_i \sigma^{z}_{i+1})} e^{(-i (\frac{\pi}{2} -\epsilon)\sum_i \sigma^{x}_i)}.
\label{propagator}
\ee
When $\epsilon = 0$, the spin chain trivially exhibits time correlations at twice the driving frequency for any initial state that spontaneously breaks the $\mathbb{Z}_2$ symmetry of $H_0$.  However, in order to qualify as a DTC, there should be a class of initial states for which some physical observables must exhibit indefinitely long sub-harmonic oscillations in the thermodynamic limit, even when $\epsilon \ne 0$ \cite{fazio2017prb,liu2018prl}. This kind of TTSB occurs in our model, when the Ising interaction is non-zero \cite{jaksch2019pra}.  \\

In this paper, we study the response of the spin chain to the periodic drive by computing the return probability:
\be 
P(t) = \vert \langle \psi(t) \vert \psi(0) \rangle \vert^2
\label{le}
\ee 
If $P(t)$ exhibits robust sub-harmonic oscillations for long times, we conclude that the system is in the DTC phase. In the DTC phase, the stroboscopic return probability at times $2nT$ remains almost constant for a very long time, and the DTC lifetime, $n^{*}$ is usually defined to be the number of drive periods, after which $P(2nT)$ falls below a critical value ($\sim 0.05$) \cite{liu2018prl,zhu2019njp,rossini2020pra}. The rationale for using this measure to quantify lifetime is the following: the Fourier transform of $P(t)$ taken up to a time $t < 2n^{*}T$ exhibits a peak at $\omega/2$, whereas this peak splits when the Fourier transform is taken over longer times; thus, the DTC exhibits persistent sub-harmonic oscillations at a rigid rhythm up to a time, $t \sim 2n^{*} T$. In the next section, we systematically explore the dependence of the DTC lifetime on the Ising interaction strength.

\section{Maximizing the DTC lifetime}

\begin{figure*}
\includegraphics[scale=0.25]{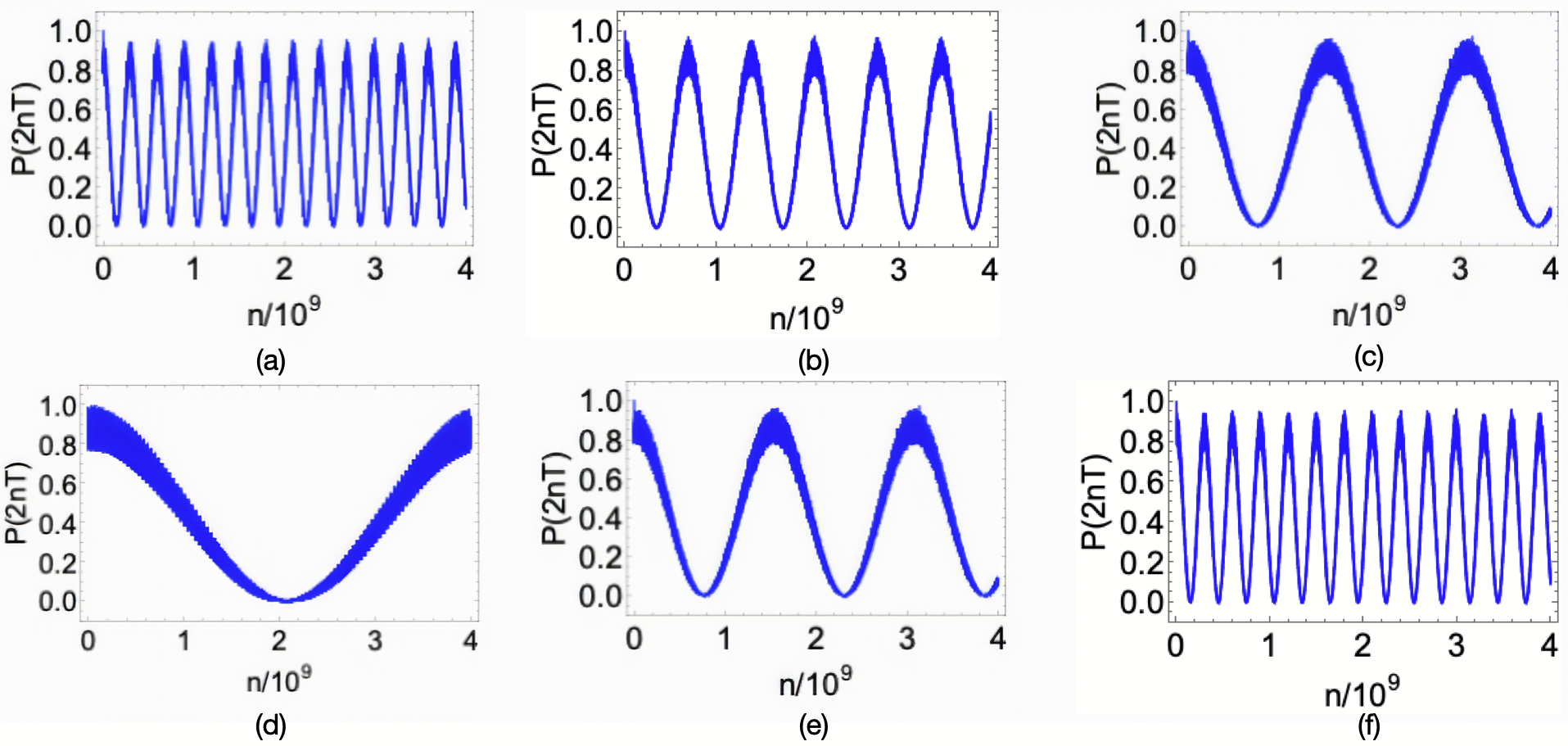}
\caption{
(a)-(f): The strobosocopic return probability at times $2nT (P(2nT))$ for a 11 site spin chain with nearest-neighbor interactions; $\frac{JT}{\pi}=0.75$, $0.85$, $0.95$, $1$, $1.05$, and $1.25$ respectively, and $\epsilon=0.1\pi$. We find that time crystal signatures can be observed for a wide range of $JT$, and the time crystal lifetime is maximum when $JT = \pi$. }
\label{fig4}
\end{figure*}

In this section, we study the stroboscopic evolution of a $L$-site periodic chain, prepared in a spin-polarized initial state i.e. $\vert \psi (t=0) \rangle$ can be either $\vert  \ldots \uparrow \uparrow \uparrow \ldots \rangle$ or $\vert \ldots \downarrow \downarrow \downarrow  \ldots \rangle$ . We analyze the response of the system to imperfect $\pi-$pulses in the experimentally relevant parameter regime, where $\sin(\epsilon) \approx \epsilon$ and $\cos(\epsilon) \approx 1-\frac{\epsilon^2}{2}$.  

\subsection{Few-cycles dynamics }

To the lowest order in $\epsilon$, we obtain the following expression for the return probability after the first two pulses:
\bea
P(2 T) &=& |\cos(\epsilon)^{(2 L)} - L \exp(-i J T) \cos(\epsilon)^{(2 L-2)} \sin(\epsilon)^2|^2 \nonumber \\
&\approx&  |(1-\frac{\epsilon^2}{2})^{(2 L)} - L\exp(-i J T) (1-\frac{\epsilon^2}{2})^{(2 L-2)} (\epsilon)^2|^2 \nonumber \\
&\approx& \vert 1 - L \epsilon^2 \left(1+\exp(-i J T)\right) \vert^2.
\eea
It is straightforward to see, that in this limit, an almost perfect revival of the initial state occurs when $JT = \pi$. This revival implies that the system can exhibit persistent oscillations of physical order parameters (like the magnetization) with a frequency $\frac{\pi}{T}$ - a direct signature of discrete TTSB. \\

This remarkable conclusion can be understood by a very simple physical picture. Let us assume that the system is initially prepared in a spin-polarized state: $\vert \psi(t=0)\rangle =\vert  \ldots \uparrow \uparrow \uparrow \ldots \rangle$. Thus, after one imperfect $\pi$ pulse, the state of the system is:
\bea
\vert \psi(t=T^{+}) \rangle &\approx& \cos(\epsilon)^L \vert  \ldots \downarrow \downarrow \downarrow \ldots \rangle - \cos(\epsilon)^{L-1} \sin(\epsilon)\nonumber \\
& &  \sum_h \vert  \ldots \downarrow \ldots \downarrow \uparrow_h \downarrow \ldots \downarrow \ldots \rangle
\eea
Just before the second $\pi$ pulse is applied, the spin chain wavefunction is :
\bea
\vert \psi(t=2 T^{-}) \rangle &\approx& \cos(\epsilon)^L \vert  \ldots \downarrow \downarrow \downarrow \ldots \rangle - \cos(\epsilon)^{L-1} \sin(\epsilon) \nonumber \\
& e^{-i JT} & \sum_h \vert  \ldots \downarrow \ldots \downarrow \uparrow_h \downarrow \ldots \downarrow \ldots \rangle 
\eea
Now, after the second $\pi$ pulse is applied the  state of the system becomes:
\bea
\vert \psi(t=2 T^{+}) \rangle &=&  (\cos(\epsilon)^{(2 L)} - L e^{(-i J T)} \cos(\epsilon)^{(2 L-2)} \sin(\epsilon)^2)  \nonumber \\
&& \vert  \psi(t=0) \rangle + \ldots
\eea
Thus when the condition $JT = \pi$ is satisfied,  $\vert \psi(t=2 T^{+}) \rangle \approx\vert \psi(t=0)\rangle$. In other words, a revival of the initial state occurs after two pulses due to a constructive quantum interference. A schematic derivation of this result is illustrated in Fig.~\ref{fig2}. Thus an interaction induced quantum interference can enable the creation of a DTC - a phenomena similar to the recently proposed many-body echo \cite{chin2019nature,zhai2019arxiv}. A similar scheme has been studied for the infinite range interacting Ising model \cite{zhou2019arxiv}; in that case $JT=\pi$ leads to a perfect revival of the initial state for even size chains, while there is no DTC order for odd size chains. However, for the nearest-neighbor Ising model, $JT=\pi$ is the optimal interaction strength for both odd and even size chains.

\subsection{Long time dynamics}
A perfect revival of the initial state would lead to an eternal time crystal. However, in our model, the revival is imperfect, and the discrete time crystalline order may disappear at later times. In order to understand what happens to the DTC at later times, we first study the dynamics of the spin chain in the few-cycles regime. In this case, when $\epsilon$ is small, the evolution of the spin chain can be captured by an effective analytic model  \cite{jaksch2019pra}. This effective model is obtained by truncating the Hilbert space to only two kinds of states: (a) Fully polarized states, and (b) One magnon states i.e. states with $(L-1)$  spin-ups (spin-downs), and 1 spin-down (spin-up). Using this effective model, we find that the spin chain wavefunction after $2n$ pulses is given by:
\be
\vert \psi (2nT) \rangle = c_0 (2 nT) \vert \psi_0 \rangle + c_1 (2 nT)  \sum_h \vert h \rangle,
\ee
where $\vert \psi_0 \rangle = \vert  \ldots \uparrow \uparrow \uparrow \ldots \rangle $ (the fully polarized state), $\vert h\rangle = \vert  \ldots \uparrow \ldots \uparrow \downarrow_h \uparrow \ldots \uparrow \ldots \rangle$ (the one-magnon states), and 
\be 
c_1 (2 nT) = \epsilon \sum_{j=1}^{2nT-1} \exp(-i j JT).
\ee
This implies that 
\be
\vert c_1 (2 nT) \vert = \epsilon \vert \frac{\sin(nJT)}{\sin(TJ/(2))} \vert,
\ee
and thus the stroboscopic return probability $P(2nT)$ is given by:
\be
P(2nT) = c_0 (2 nT)^2 = 1- L \epsilon^2 \left(\frac{\sin(nJT)}{\sin(TJ/(2))} \right)^2.
\ee
When $JT=0$, the return probability decays very fast, implying the absence of any TTSB. In the presence of interactions however, the spins can become synchronized and the time crystal order can persist. Furthermore,  when $JT = \pi$, $P(2nT) \approx 1$, thereby indicating that the system may exhibit time crystalline behavior for long times. \\

\begin{figure*}
\includegraphics[scale=0.25]{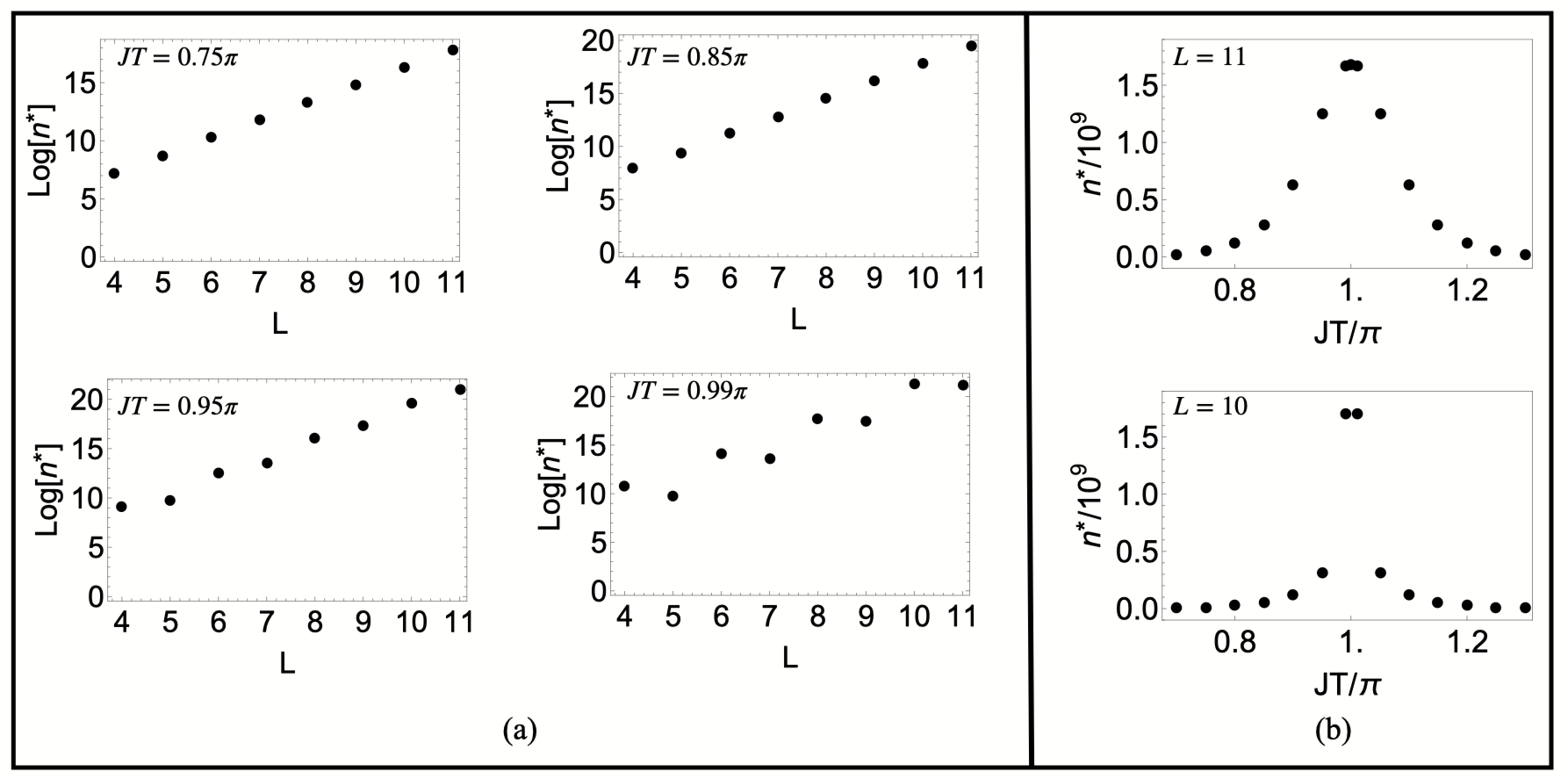}
\caption{(a) The dependence of the DTC lifetime, $n^{*}$ defined below Eq.~(\ref{le}) on the system size for various values of $JT$. We find that the lifetime increases exponentially with the system size. When $JT \approx \pi$, the spin chain exhibits an interesting feature: the lifetime of an even size system of length $L$ is almost the same as that of an odd size system of length $L+1$. This is due to a constructive interference from several paths for even size chains. (b) The dependence of the lifetime on $JT$ for a fixed spin chain length, $L$. It is evident that $n^{*}$ is maximum, when $JT = \pi$ for both even and odd size chains. }
\label{fig5}
\end{figure*}

\begin{figure*}
\includegraphics[scale=0.25]{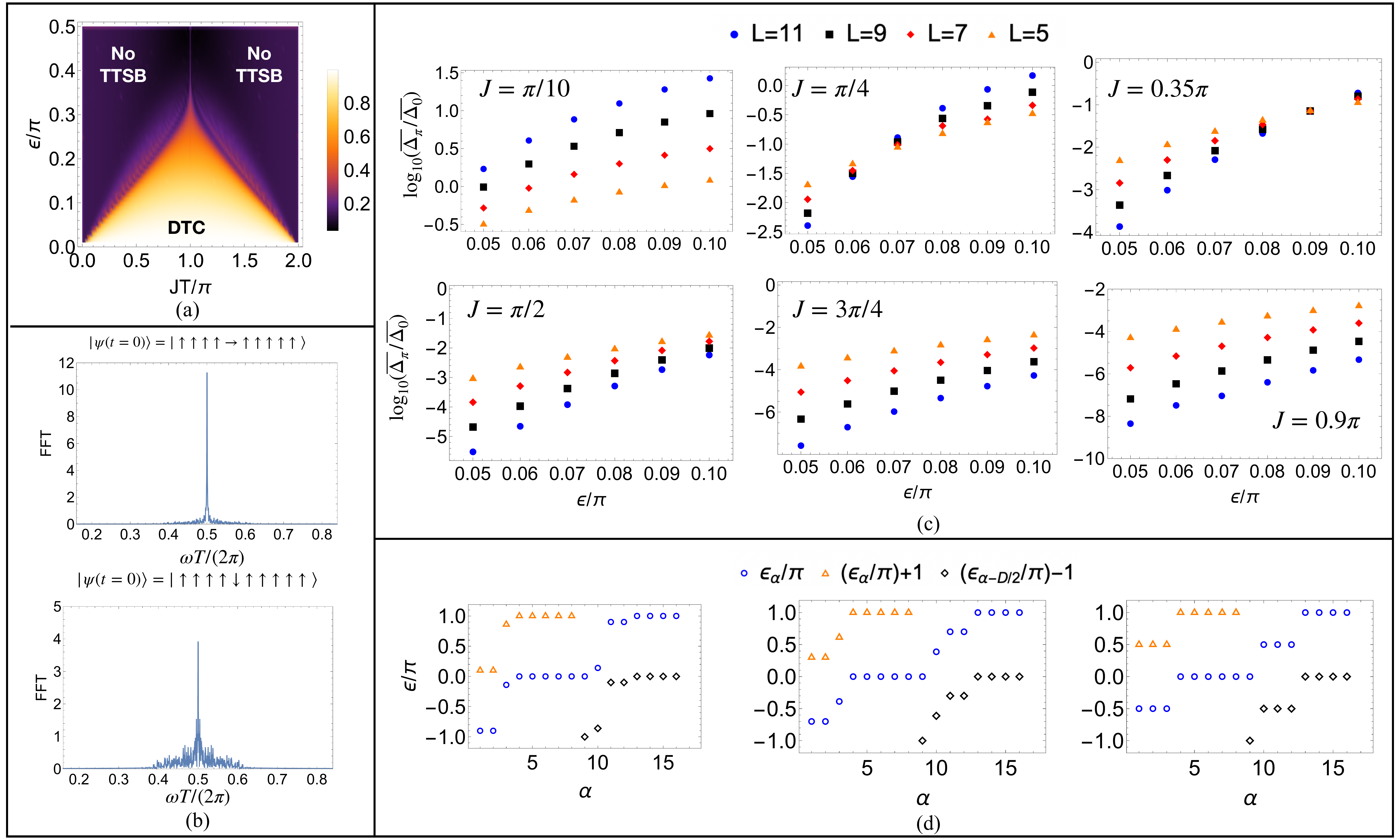}
\caption{ (a) Phase diagram of the model obtained from averaging $P(2nT)$ over 1000 periods for a 10-site system. There is a large region of parameter space where the system exhibits DTC order. (b) The Fourier spectrum of $\sigma_i^z(t)$ when the $i$-th spin is polarized along the $+x$ and $-z$ direction respectively, while the other spins are polarized along $+z$ direction. It is clear that the system exhibits a sharp Fourier peak at $\omega = \pi/T$ - a direct signature of period doubling oscillations. (c) The eigenstate averaged energy gaps defined in Eq.~\ref{gap1} and Eq.~\ref{gap2} as a function of $\epsilon$ and $J$ for different system sizes $L$. In the DTC regime, $\frac{\Delta_{\pi}}{\Delta_0} \rightarrow 0$ in the thermodynamic limit. It is evident from this figure that the transition from the no-TTSB phase to the DTC phase is accompanied by a transition from increasing $\frac{\Delta_{\pi}}{\Delta_0}$ to decreasing $\frac{\Delta_{\pi}}{\Delta_0}$ with increasing system size. (d) Quasi-energy spectrum for a 4-site system when $JT=\pi$: there are 4 perfect $\pi$-spectrally paired states due to the time reflection symmetry that emerges at this point. These perfect $\pi$-spectral pairs would lead to eternal period doubling oscillations.}
\label{fig6}
\end{figure*}

Next, we investigate the fate of the system at even longer times, by performing exact diagonalization calculations on a 11 site model, when $\epsilon = 0.1 \pi$.  Our results are shown in Fig.~\ref{fig4}. It is evident from this figure that while there is a large parameter regime, where the spin chain exhibits sub-harmonic oscillations at a frequency of $\pi/T$, the DTC lifetime is maximized when $JT=\pi$.  To further substantiate our finding, we analyze the dependence of the DTC lifetime, $n^{*}$ on system size for some fixed values of $JT$.  As shown in Fig.~\ref{fig5}(a), we find that $n^{*}$ increases exponentially with the system size. This is a characteristic feature of DTCs. Interestingly, we find that when $JT\approx \pi$, the lifetime of even size chains of length $L$ can be greater that of odd size chains of length $L+1$.  As we shall explain below, this behavior can be understood as a consequence of a time reflection symmetry of the model at $JT= \pi$. It is also worth noting that the behavior of $P(2nT)$ is symmetric around $JT=\pi$, thereby implying that $n^{*}$ is the same for $JT=y$ and $JT = 2 \pi - y$, where $y \in \mathbb{R}$. We provide numerical evidence for this result in Fig.~\ref{fig5}(b). The results from Fig.~\ref{fig5} suggests the existence of a critical $J$, below which TTSB would not occur for every value of $\epsilon$. We proceed to determine the region of parameter space, where the system is a DTC by evaluating the average stroboscopic return probability, $\overline{P(2nT)}$ over $1000$ periods. Our results are shown in Fig.~\ref{fig6}(b).\\

Our results so far have focused on the period doubling oscillations exhibited by the spin chain, when it is initialized in a globally polarized state. While these oscillations suggest that the system may be a DTC, in order to rigorously establish the existence of the DTC phase, we have to examine whether the model possesses two crucial characteristics: (a) it has to be integrable, and (b) it must host an extensively large number of eigenstate pairs with quasi-energies differing by $\pi/T$. The integrability of this system is already well established \cite{prosen2000exact,prosen2002general}; this feature allows it to evade thermalization. In order to check for the ``$\pi$-spectral pairing", we compute the energy difference between neighboring eigenstates:

\begin{equation}
\label{gap1}
    \Delta_0 =\epsilon_{i+1}-\epsilon_i,
\end{equation} 
as well as 
\begin{equation}
   \Delta_{\pi}= =\epsilon_{i+D/2}-\epsilon_i-\pi/T, 
\label{gap2}
\end{equation}
where $\epsilon$ is the quasi-energy and $D=2^L$ is the Hilbert Space dimensions. Robust ``$\pi$-spectral pairing" occurs, when $\Delta_{\pi} \ll \Delta_{0}$ \cite{sondhi2016prb}. Our results are shown in Fig.~\ref{fig6}(c); it is quite clear that the model hosts an extensive number of $\pi$-spectral paired eigenstates in the parameter regime that we had earlier identified to be a DTC. Consequently, we find that the system can exhibit period doubling oscillations for different initial states (see Fig.~\ref{fig6}(b)). This analysis of the quasi-energy spectrum clearly establishes this model as a DTC. \\

Finally, we address the ``odd-even" effect in the DTC lifetime, when $JT = \pi$. Intriguingly at this special point, the system possesses a time reflection symmetry generated by the operator \cite{iadecola2018floquet}:
\begin{equation}
    R = \prod_{i=1}^{L} \sigma_i^x \prod_{j=1}^{L} \sigma_j^z,
\end{equation}
such that
\begin{equation}
    R U(T) R^{\dagger} = \exp(- i L \pi/2) U(T),
\end{equation}
where $U(T)$ is the Floquet propagator defined in Eq.~\ref{propagator}.\\

As a consequence of this symmetry, there are at least $2^{L/2}$ states exactly at quasi-energies $0$ and $\pi$ for even size systems (for a detailed derivation of this result, see ref.~\cite{iadecola2018floquet}). This result implies that there are at least $2^{L/2}$ exact $\pi$-spectral paired eigenstates even in a finite size system (see Fig.~\ref{fig6}(d)). We note that generically such exact $\pi$-spectral pairing occurs only in thermodynamic limit. The presence of these exact $\pi$-spectral pairs has a striking consequence: the spin chain can exhibit eternal period doubling oscillations. To see this, let us assume that the system is initially prepared in a state:
\begin{equation}
    \vert \psi (t=0) \rangle = \vert \overline{\psi}_{\pm} \rangle = \vert \phi_0 \rangle \pm \vert \phi_{\pi} \rangle,
\end{equation}. 
where $\phi_0 \rangle$ ($\phi_{\pi} \rangle$ is a Floquet eigenstate with quasi-energy $0(\pi)$. In this case:
\begin{equation}
    \vert \psi (t=T) \rangle = \vert \overline{\psi}_{\mp} \rangle = \vert \phi_0 \rangle \mp \vert \phi_{\pi} \rangle
\end{equation} and 
\begin{equation}
     \vert \psi (t=2T) \rangle = \vert \overline{\psi}_{\pm} \rangle = \vert \phi_0 \rangle \pm \vert \phi_{\pi} \rangle = \vert \psi (t=0) \rangle
\end{equation}
This perfect revival after two pulses implies that the stroboscopic return probability, $P(2nT)$ is always $1$ and the DTC has eternal lifetime. Of course the states $\vert \overline{\psi}_{\pm} \rangle $ is generally difficult to prepare experimentally. However, these eternal oscillations can be observed as long as the initial state has a significant overlap with $\vert \overline{\psi}_{\pm} \rangle$. For the finite size systems considered here, the globally polarized state is indeed an example of these special class of initial states, and therefore the chain behaves as an eternal DTC, when it is initialized in this state. Furthermore, the lifetime of the DTC in even size systems is significantly enhanced when $\vert JT-\pi\vert \ll 1$.\\

A natural question to investigate at this point is the effect of disorder on this system, when $JT \approx \pi$. By comparing to known results in the literature, we conclude that the presence of strong onsite disorder would destroy the DTC in this parameter regime \cite{yao2017prl}. A drawback of our scheme is that an eternal DTC can only be realized for a limited class of initial states. This drawback can be overcome by engineering infinite range interactions \cite{zhou2019arxiv}.\\

\subsection{Experimental Realization}
Before we conclude, it is instructive to discuss potential experimental realizations of our proposal. A promising platform to test our predictions is a Rydberg atom chain \cite{rossini2020pra}. In this system, the atomic ground state and a chosen Rydberg state can be mapped to a pseudospin-1/2 system; the periodic kick can be simulated using a periodically modulated laser beam that couples these states \cite{guardado2018probing,labuhn2016tunable,schauss2018quantum}.  Alternatively, this model can be engineered using ultracold bosons loaded in a tilted optical lattice in the Mott insulator regime \cite{simon2011quantum}. Finally, we note that digital quantum simulations on superconducting quantum processors \cite{ippoliti2020many,kyriienko2018floquet} and trapped ion quantum computers \cite{lanyon2011universal,monroe2019programmable} provide an alternative route to realize the long lived DTC we have studied in this paper. 

\section{Summary and Outlook}
A time crystal is an intriguing non-equilibrium phase of matter, that is expected to be a very useful platform for performing precision measurements \cite{richermephysics,gibney2017nature} and quantum simulation \cite{estarellas2020simulating}. In this paper, we have described a scheme to extend the lifetime of a DTC in a periodically driven finite Ising spin chain with nearest-neighbor interactions and no disorder. Our major insight is that the lifetime of the DTC can be maximized by tuning the interaction strength to an optimal value. This is a consequence of a novel interaction induced quantum interference mechanism.  Furthermore, we find that the DTC lifetime grows exponentially with the system size, with a considerably greater enhancement for even size chains at the optimal interaction strength. Finally, we have discussed possible realizations of our model in various quantum simulator platforms.\\

One of the grand challenges of Floquet engineering is to determine an optimal frequency window, where heating is suppressed \cite{eckardt2019prr}. Our scheme can overcome problems associated with high-frequency driving, and thus it can play an important role in designing future experiments on time crystals. While we have considered the case of nearest neighbor interactions in this paper, it will be interesting to extend our treatment to long range interacting systems in the future. In an exciting development, Viebahn {\it et al.} have recently demonstrated that a similar two path interference can be employed to suppress heating in a periodically driven ultracold fermionic system \cite{esslinger2020arxiv}. This leads us to believe that generalizations of our scheme may be useful for stabilizing other Floquet systems, and we plan to explore this in future work. 

\acknowledgments
The author thanks Qi Zhou for discussions and encouragement, and the Wilczek Quantum Center for its hospitality while part of this work was performed. This work is supported by the AFOSR Grant No. FA9550-16-1-0006, and the MURI-ARO Grant No. W911NF17-1-0323 through UC Santa Barbara.\\

\bibliography{ref} 

\end{document}